\documentclass[apj]{emulateapj}
\usepackage{epstopdf}
\usepackage{longtable}
\usepackage{fancyvrb}
\usepackage{multirow}
\usepackage{verbatim}
\usepackage{longtable}
\usepackage[caption=false]{subfig}
\usepackage{float}
\usepackage{enumerate}
\newcommand{\z}{$z$}

\newcommand{\oiii}{[\ion{O}{3}]$\lambda5007$}

\newcommand{\LOIII}{$L_{\rm{[OIII],corr.}}$}

\newcommand{\LXHAGN}{$L_{\rm{2-10keV,unabs.}}$}

\newcommand{\LXU}{$L_{\rm{14-195keV}}$}

\newcommand{\hr}{$HR$}

\newcommand{\deltatheta}{$\Delta \Theta$}
\newcommand{\deltathetasig}{$\sigma_{\Delta \Theta}$}
\newcommand{\deltathetasim}{$\Delta \Theta_{\rm{sim.}}$}
\newcommand{\deltathetasigsim}{$\sigma_{\Delta \Theta,\rm{sim.}}$}

\newcommand{\deltas}{$\Delta S_{\rm{proj.}}$}
\newcommand{\deltassim}{$\Delta S_{\rm{proj.,sim.}}$}
\newcommand{\deltaspot}{$\Delta S_{\rm{proj.,pot.}}$}
\newcommand{\zsim}{$z_{\rm{sim.}}$}

\newcommand{\pnull}{$p_{\rm{null}}$}

\newcommand{\ngroup}{$N_{\rm{group}}$}

\newcommand{\lbol}{$L_{\rm{Bol}}$}

\newcommand{\infib}{\texttt{In-Fiber}}
\newcommand{\outfib}{\texttt{Out-Fiber}}

\newcommand{\parentsz}{48}
\newcommand{\parentnobiassz}{40}

\newcommand{\catabAGNsz}{18}

\newcommand{\offbiassz}{8}
\newcommand{\offnobiassz}{10}

\newcommand{\dualsdsssyssz}{3}
\newcommand{\dualsdsssysszlet}{three}
\newcommand{\dualsdsssz}{6}
\newcommand{\parentsdsssz}{69}

\newcommand{\dualbatsyssz}{8}
\newcommand{\dualbatsz}{16}
\newcommand{\parentbatsz}{246}

\newcommand{\comerforddualsysszlet}{one}

\newcommand{\liudualsysszlet}{two}

\newcommand{\binmin}{2}

\newcommand{\offsetLBol}{$0.04\sigma$} 
\newcommand{\offsetLBolnocorr}{$0.02\sigma$} 
\newcommand{\dualsdssLBol}{$3.38\sigma$} 
\newcommand{\dualbatLBol}{$1.82\sigma$} 
\newcommand{\offsetPhySep}{$2.74\sigma$} 
\newcommand{\offsetPhySepnocorr}{$2.41\sigma$} 
\newcommand{\dualsdssPhySep}{$2.61\sigma$} 
\newcommand{\dualbatPhySep}{$1.46\sigma$} 
\newcommand{\offsetLBolA}{$1.44_{-1.42}^{+185}$} 
\newcommand{\offsetLBolB}{$9.98_{-216}^{+207}\times 10^{-4}$} 
\newcommand{\offsetLBolAnocorr}{$1.21_{-1.20}^{+175}$} 
\newcommand{\offsetLBolBnocorr}{$3.25_{-153}^{+142}\times 10^{-4}$} 
\newcommand{\dualsdssLBolA}{$3.93_{-3.46}^{+3.66}\times 10^{-4}$} 
\newcommand{\dualsdssLBolB}{$7.74_{-2.28}^{+2.06}\times 10^{-2}$} 
\newcommand{\dualbatLBolA}{$4.26_{-1.63}^{+2.58}\times 10^{-1}$} 
\newcommand{\dualbatLBolB}{$8.68_{-4.78}^{+4.99}\times 10^{-3}$} 
\newcommand{\offsetPhySepA}{$2.04_{-0.10}^{+0.10}\times 10^{-1}$} 
\newcommand{\offsetPhySepB}{$-1.28_{-0.43}^{+0.46}$} 
\newcommand{\offsetPhySepAnocorr}{$1.63_{-0.08}^{+0.08}\times 10^{-1}$} 
\newcommand{\offsetPhySepBnocorr}{$-1.21_{-0.51}^{+0.50}$} 
\newcommand{\dualsdssPhySepA}{$1.09_{-0.03}^{+0.03}$} 
\newcommand{\dualsdssPhySepB}{$-3.33_{-1.34}^{+1.38}\times 10^{-2}$} 
\newcommand{\dualbatPhySepA}{$1.66_{-0.39}^{+0.39}$} 
\newcommand{\dualbatPhySepB}{$-1.50_{-0.93}^{+1.02}\times 10^{-1}$} 

\newcommand{\wise}{\emph{WISE}}
\newcommand{\ch}{\emph{Chandra}}

\newcommand{\full}{\emph{complete}}
\newcommand{\sdss}{SDSS}
\newcommand{\bat}{BAT}

\newcommand{\offsetfrac}{$f_{\rm{Offset}}$}
\newcommand{\dualfracsdss}{$f_{\rm{Dual,SDSS}}$}
\newcommand{\dualfracbat}{$f_{\rm{Dual,BAT}}$}
\newcommand{\noff}{$n_{\rm{Offset}}$}
\newcommand{\npar}{$n_{\rm{Parent}}$}
\newcommand{\nparcum}{$n_{\rm{Parent[\geq\Delta S]}}$}

\newcommand{\offa}{SDSSJ111458.02+403611.41}
\newcommand{\offashort}{SDSSJ1114+4036}
\newcommand{\offashortone}{SDSSJ1114+4036NE}
\newcommand{\offashorttwo}{SDSSJ1114+4036SW}
\newcommand{\duala}{SDSSJ110851.04+065901.5}
\newcommand{\dualashort}{SDSSJ1108+0659}
\newcommand{\dualashortone}{SDSSJ1108+0659NW}
\newcommand{\dualashorttwo}{SDSSJ1108+0659SE}
\newcommand{\dualb}{SDSSJ114642.47+511029.6}
\newcommand{\dualbshort}{SDSSJ1146+5110}
\newcommand{\dualbshortone}{SDSSJ1146+5110SW}
\newcommand{\dualbshorttwo}{SDSSJ1146+5110NE}
\newcommand{\dualc}{SDSSJ112659.54+294442.8}
\newcommand{\dualcshort}{SDSSJ1126+2944}
\newcommand{\dualcshortone}{SDSSJ1126+2944NW}
\newcommand{\dualcshorttwo}{SDSSJ1126+2944SE}

\newcommand{\paperI}{Paper I}

\shorttitle{Spatially Offset AGN II}
\shortauthors{Barrows et al.}

\begin{document}

\submitted{Accepted for publication in ApJ}

\title{Spatially Offset Active Galactic Nuclei. II: Triggering in Galaxy Mergers}
\author{R. Scott Barrows$^{1}$, Julia M. Comerford$^{1}$, Jenny E. Greene$^{2}$, and David Pooley$^{3}$}
\address{$^{1}$Department of Astrophysical and Planetary Sciences, University of Colorado Boulder, Boulder, CO 80309, USA; Robert.Barrows@Colorado.edu}
\address{$^{2}$Department of Astrophysical Sciences, Princeton University, Princeton, NJ 08544, USA}
\address{$^{3}$Department of Physics and Astronomy, Trinity University, San Antonio, TX 78212, USA}

\bibliographystyle{apj}

\begin{abstract}

Galaxy mergers are likely to play a role in triggering active galactic nuclei (AGN), but the conditions under which this process occurs are poorly understood.  In \paperI, we constructed a sample of spatially offset X-ray AGN that represent galaxy mergers hosting a single AGN.  In this paper, we use our offset AGN sample to constrain the parameters that affect AGN observability in galaxy mergers.  We also construct dual AGN samples with similar selection properties for comparison.  We find that the offset AGN fraction shows no evidence for a dependence on AGN luminosity, while the dual AGN fractions show stronger evidence for a positive dependence, suggesting that the merger events forming dual AGN are more efficient at instigating accretion onto supermassive black holes than those forming offset AGN.  We also find that the offset and dual AGN fractions both have a negative dependence on nuclear separation and are similar in value at small physical scales.  This dependence may become stronger when restricted to high AGN luminosities, though a larger sample is needed for confirmation.  These results indicate that the probability of AGN triggering increases at later merger stages.  This study is the first to systematically probe down to nuclear separations of $<1$ kpc ($\sim0.8$ kpc) and is consistent with predictions from simulations that AGN observability peaks in this regime.  We also find that the offset AGN are not preferentially obscured compared to the parent AGN sample, suggesting that our selection may be targeting galaxy mergers with relatively dust-free nuclear regions.

\end{abstract}


\section{Introduction}
\label{sec:intro}

Accretion onto supermassive black holes (SMBHs), and the corresponding release of gravitational potential energy, power active galactic nuclei (AGN).  This process requires that a significant amount of matter in the interstellar medium of a galaxy experience a loss of angular momentum sufficient to ultimately be captured by the SMBH's accretion disk.  

Numerical simulations suggest that major mergers of galaxies are an effective mechanism for removing angular momentum \citep{Barnes:1991,Mihos:1996,Hopkins05,Springel2005}.  Observational evidence for this scenario includes bright quasi-stellar objects (QSOs) that often show evidence of interactions or mergers at both obscured phases, such as dust-reddened QSOs \citep{Glikman:2015} and ultra-luminous infrared galaxies \citep[ULIRGS; ][]{Sanders:1988,Sanders:1988a,Canalizo:2001}, and in traditional QSOs \citep{Hong:2015}.  This scenario is also consistent with the hierarchical paradigm of galaxy evolution, in which massive stellar bulges are capable of fueling SMBH growth through mergers of gas rich galaxies \citep{Hopkins2008,Younger:2008} and suggests that SMBH growth may be linked with merger events in order to maintain observed correlations between SMBHs and their host galaxies.  For example, the masses of SMBHs appear to be correlated with the stellar velocity dispersion \citep{Gebhardt00,Ferrarese2000,Gultekin:2009} and luminosities \citep{Marconi:Hunt:2003,Bentz:2009c} of the central stellar bulges, implying that the buildup of SMBH mass is correlated with the buildup of stellar bulge mass \citep{McLure:2002,Haring:2004}.

The structures of disk galaxies, on the other hand, are thought to be shaped by the collapse of gas via energy dissipation and smooth accretion from cooling gas within the surrounding dark matter halo, with a previous phase of mergers shaping the central stellar bulge and stellar halo \citep{Blumenthal:1984,Debattista:2006,Robertson:2006}.  While subsequent major mergers will not necessarily destroy the disk if the gas supply of the progenitors is very high and the resulting bulge component is small \citep{Springel:Hernquist:2005,Hopkins:2009b}, simulations have been successful at reproducing the observed properties of more massive bulges (corresponding to substantial SMBH growth) through major mergers of spiral galaxies \citep{Toomre:1977,DiMatteo:2005,Cox:2006}, implying that AGN hosted by disk galaxies may be triggered via alternative routes.  Such mechanisms include instabilities internal to the host galaxy \citep{Lynden-Bell:1979,Sellwood:1981,VanAldaba:1981,Combes:1985,Pfenniger:1991,Heller:1994,Bournaud:2002,Athanassoula:2003,Sakamoto:1999} and minor mergers \citep{Taniguchi:1999,Corbin:2000}.

However, the relative roles of internal instabilities, minor mergers and major mergers for growing SMBHs is unclear.  For example, the majority of local AGN reside in late-type galaxies unlikely to have experienced a major merger \citep{Dong:DeRoberties:2006}, though their luminosities do not extend to the more powerful regimes seen in high-redshift quasars.  These observables are explained in the model of \citet{Hopkins:Hernquist:2006} where secular SMBH growth dominates in the local Universe but remains important only for low-luminosity AGN triggering toward high-redshifts as the rate of gas-rich mergers increases.  However, finding direct and consistent evidence in support of this picture has been difficult.  Numerous studies find that high luminosity AGN show no preference for existing in galaxies with signs of merger activity when they are compared to a control sample of inactive galaxies (\citealp{Georgakakis:2009,Kocevski:2012,Simmons:2012,Villforth:2014,Mechtley:2015,Villforth:2016}, though see \citealt{Schawinski:2012} for a potential luminosity dependence).  On the other hand, several studies find that the fraction of AGN in mergers, out of a parent AGN sample, increases with increasing luminosity \citep{Treister:2012,Comerford:Greene:2014,Glikman:2015}.

The above qualitative disagreements may be due to the variable conditions under which AGN triggering may happen within a galaxy merger.  For example, the stage of the merging galaxy system may be a particularly important parameter for the triggered accretion rate.  The tidal torques induced by the merger are predicted to funnel gas and dust toward the nuclear region of the merging system, and as the merger evolves the SMBHs will lose angular momentum due to dynamical friction, thereby migrating toward the nuclear region as well.  Therefore, at later merger stages, and thus smaller SMBH separations, the supply of gas for accretion is greater, such that the accretion rate is likely to be higher \citep{Van_Wassenhove:2012,Blecha:2013}, albeit with considerable uncertainty as to the timescales of AGN duty cycles.  While several studies have examined the dependence on merger stage using nuclear separations as a proxy \citep{Ellison:2011,Silverman:2011,Koss:2012}, spatial resolution limits have precluded systematic analyses from observing the small nuclear separations when the accretion rate is predicted by simulations to peak \citep{Stickley:2014}.  Furthermore, simulations predict that the probability of triggering one versus two AGN within a merger may be different and depend on properties of the host galaxies, such as the mass of their nuclear stellar cores \citep{Yu:2011,Capelo:2015}.  Additionally, nuclear obscuration in mergers may hinder the establishment of a connection between mergers and AGN at optical wavelengths.  Indeed, \citet{Kocevski:2015} found that the fraction of galaxies with disturbed morphologies increases with the level of nuclear obscuration, suggesting that this may be a key phase in the evolution of AGN in galaxies but which is hidden from most observations.

To understand the conditions under which AGN triggering is correlated with galaxy mergers, uniform merger samples with well-understood selection biases are necessary.  While many galaxy merger candidates have been selected spectroscopically from velocity offset AGN emission lines \citep{Comerford2009a,Wang2009,Liu2010a,Barrows:2012,Ge:2012,Barrows:2013}, the majority of them have been shown to host AGN-driven outflows rather than dual SMBHs based on follow-up observations \citep{Mueller-Sanchez:2016,Nevin:2016}.  From imaging, several samples have been selected based on morphology, either visually \citep{Kocevski:2012} or based on asymmetry \citep{Villforth:2014}.  However, selection by morphology is not necessarily capable of quantifying the merger stage accurately if two nuclei are not visible.  While samples based on galaxy pairs can measure separations, they are necessarily limited to earlier merger stages \citep{Iwasawa:2011,Ellison:2011,Silverman:2011,Liu:2012a,Satyapal:2014}.  

However, resolved separations can be significantly reduced by spatially constraining the relative locations of two individual SMBHs in a merging system \citep{Lackner:2014,Mueller-Sanchez:2015}.  In this study, we exploit this concept by using a newly constructed sample of X-ray AGN that are spatially offset from the nucleus of the host galaxy or a nearby companion galaxy from \citet{Barrows:2016}, hereafter referred to as \paperI.  The spatially offset AGN sample can be used to quantify the merger stage based on separation, and was constructed using an astrometric registration procedure that detects offsets down to $<1$ kpc.  Additionally, the sample also allows us to compare the merger scenarios for single AGN formation against those of dual AGN formation.  This paper is organized as follows: in Section \ref{sec:sample} we described our sample, in Section \ref{sec:analysis} we analyze the effects of bolometric luminosity, nuclear separation, group environment and obscuration on the samples, in Section \ref{sec:discussion} we discuss the conditions that affect AGN triggering in galaxy mergers and Section \ref{sec:conclusions} contains our conclusions.  We assume the cosmological parameters of $H_{0}=70$ km s$^{-1}$ Mpc$^{-1}$, $\Omega_{m}=0.3$ and $\Omega_{\Lambda}=0.7$ throughout.

\section{Samples}
\label{sec:sample}

In this section, we discuss the samples used in our analysis: an offset AGN sample and two separate dual AGN samples.  The offset AGN represent galaxy mergers hosting only a single AGN that is off-nuclear (Section \ref{sec:offset_agn}), and the dual AGN represent galaxy mergers hosting two AGN (Section \ref{sec:dual_agn}).

\subsection{The Offset AGN Sample}
\label{sec:offset_agn}

For our sample of galaxy mergers hosting only a single AGN, we use the spatially offset AGN from \paperI.  Full details of this sample can be found in \paperI, though here we summarize the main properties.  The AGN (Type 2) were originally selected from galaxies in the SDSS Seventh Data Release (DR7) that are located in the AGN regime of the Baldwin-Phillips-Terlivich (BPT) diagram \citep{Baldwin1981,Kewley:2006}.  From overlapping archival \ch~coverage, we identified hard X-ray sources that are required to be within the SDSS fiber and satisfy the X-ray criteria of an AGN: unabsorbed hard X-ray luminosity of \LXHAGN~$\ge10^{42}$ erg s$^{-1}$, or a hardness ratio of \hr~$\ge-0.1$ where \hr~$=(H-S)/(H+S)$ and $H$ and $S$ are the number of hard and soft X-ray counts, respectively.  The overlapping SDSS and \ch~images were astrometrically registered with our pipeline described in \paperI, establishing the \full~parent AGN sample (\parentsz).  Spatially offset X-ray AGN were then selected as those with significant spatial offsets ($\geq3\sigma$ in significance but $\leq20$ kpc in projected physical separation) from the nucleus of the host galaxy or a nearby companion galaxy, establishing the \full~offset AGN sample (\catabAGNsz).
 
Since the spatially offset AGN detections require \ch~imaging, the selection is biased toward AGN that have been targeted by \ch.  This bias will affect our analysis if the AGN were targeted as potentially being in merging systems since it will not represent a random sampling of AGN (see Section 7.2 of \paperI).  This bias affects \offbiassz~offset AGN from the \full~sample.  Removing these \offbiassz~AGN leaves a sample of \parentnobiassz~parent AGN and \offnobiassz~spatially offset AGN based on an unbiased sample of \ch~observations. These sources make up the final offset AGN sample, and they are used in all of our subsequent analyses.  Below we describe several unique characteristics of our offset AGN sample and how we address them in our analyses:

\noindent \emph{Possible Presence of Dual AGN:} Since our selection of offset AGN in \paperI~imposed several conservative X-ray thresholds, we can not rule out the possibility that any of them are dual AGN systems with the second AGN either below the X-ray source detection threshold or below the X-ray AGN thresholds (see Section 7.1 of \paperI).  One particular example is the source \duala~(\dualashort) which was selected as a spatially offset AGN in \paperI~but was also previously discussed in \citet{Liu:2013} who interpreted the system to be consistent with a dual AGN.  While \dualashort~is excluded from the sample because it was targeted by a \ch~program as a dual AGN candidate, secondary AGN that are currently undetected may exist in some of our other offset AGN systems.  For uniformity, however, we adopt the definitions used in \paperI~and note that our conclusions can only be said to apply to X-ray bright AGN.  We also acknowledge the possibility of two X-ray detected AGN in one of our sources, \offa~(\offashort; see Section 7.1 of \paperI), though we note that using the secondary X-ray source (\offashorttwo) instead of the primary X-ray source (\offashortone) has a negligible effect on our results.

\noindent \emph{SDSS Fiber Size:} As described in \paperI, due to the $1\farcs5$ SDSS fiber radius, spatial offsets between the X-ray AGN and galaxy centers may be contained entirely within the fiber or not (see Figure 3 of \paperI).  Since the X-ray AGN is always constrained to be within the fiber radius, this distinction is made between cases in which the X-ray AGN is offset from a galactic core that is inside the fiber (\infib~offset AGN) versus a galactic core that is outside the fiber (\outfib~offset AGN).  We emphasize that the distinction between the \infib~and \outfib~subsamples is not a selection effect but merely an artifact of the SDSS fiber size.  However, a side effect of this artifact is that spectroscopic coverage is available for the offset galactic cores of the \infib~subsample (smaller separation pairs with unresolved secondary nuclei in SDSS imaging) but not for the \outfib~subsample (larger separation pairs with resolved secondary nuclei in SDSS imaging).   In principle, this difference should not affect our results concerning the AGN properties assuming the X-ray AGN detection is associated with the optical AGN detection from the fiber spectrum.  Still, given that the fiber coverage of the overall systems is generally larger for the \infib~subsample, in each analysis subsection we consider the effects of removing the \outfib~offset AGN.

\noindent \emph{Nature of the X-ray Sources:} Since X-ray sources significantly in excess of \LXHAGN~$=10^{42}$ erg s$^{-1}$ are known to be associated with accreting SMBHs, the criterion of \LXHAGN~$\geq 10^{42}$ erg s$^{-1}$ is likely to rule out a non-AGN contribution \citep{Norman:2004}.  Sources that only pass the \hr~$\geq-0.1$ criterion, however, may have smaller values of \LXHAGN~and therefore more ambiguous physical origins.  The X-ray luminosity function of off-nuclear X-ray sources does not extend far above $\sim10^{41}$ erg s$^{-1}$ \citep{Sutton:2012,Mineo:2012}, with only a few brighter sources known \citep[e.g. ][]{Farrell:2009,Lin:2016}.  These objects, known as hyper luminous X-ray sources (HLXs), are often associated with intermediate mass black holes (IMBHs).  Such sources in our sample with \LXHAGN~$=10^{41}-10^{42}$ erg s$^{-1}$ are likely associated with either a lower luminosity AGN or otherwise accretion onto IMBHs with high hardness ratios \citep{Servillat:2011}, an event that is likely the result of a minor galaxy merger.  At lower luminosities (\LXHAGN~$<10^{41}$ erg s$^{-1}$), however, we can not rule out the possibility of stellar-mass systems in star-forming regions (ultra-luminous X-ray sources) that are passing through phases of unusually hard X-ray spectra that can mimic the higher hardness ratios typically seen in AGN \citep{Fabbiano:2006b,Kaaret:2009,Dewangan:2010}.  Finally, we noted in \paperI~that the ionizing nature of AGN optically classified as low-ionization nuclear emission regions (LINERs) is ambiguous and may not originate from accretion onto nuclear massive black holes \citep{Ho:1997,Komossa:1999,Terashima:2002}.  Therefore, in each analysis subsection we consider the effects of removing the subsample that does not pass the threshold of \LXHAGN~$\geq 10^{41}$ erg s$^{-1}$ or is optically classified as a LINER.

\subsection{Dual AGN Sample}
\label{sec:dual_agn}  

Our goal in this subsection is to create a sample of dual AGN from the literature for comparison to our offset AGN.  Since no samples currently exist that satisfy both the optical and X-ray selection criteria from \paperI, we have constructed two dual AGN samples, one of which is optically-selected (Section \ref{sec:sdss_dual_agn}) and the other of which is X-ray-selected (Section \ref{sec:bat_dual_agn}).

\subsubsection{The Optically-Selected Dual AGN Sample}
\label{sec:sdss_dual_agn}

Similar to the offset AGN parent sample, the parent sample of the optically-selected dual AGN is derived from the SDSS DR7 spectroscopic AGN \citep{Brinchmann:2004}.  Since we require \ch~detections to spatially isolate the AGN relative positions, we have cross-matched the SDSS spectroscopic AGN with unique detections from the \ch~Source Catalogue \citep[CSC; ][]{Evans:2010} within $1\farcs5$ to create the final parent sample of the optically-selected AGN.  From this parent sample, we use the previously identified dual AGN systems found in the two studies that have selected dual AGN starting from the SDSS spectroscopic AGN sample and for which $Chandra$ imaging reveals spatially distinct X-ray AGN \citep{Liu:2013,Comerford:2015}.  The two samples are described below:  

There are two systems from \citet{Liu:2013} that the authors classify as dual AGN: \dualashort, also in our \full~offset AGN sample (Section \ref{sec:offset_agn}), and \dualb~(\dualbshort).  The X-ray AGN in each pair are separated by $\geq3\sigma$ and $\leq20$ kpc, thereby satisfying the spatially offset criteria from \paperI.  However, in both systems one of the X-ray sources (\dualashortone~and \dualbshortone) passes the X-ray AGN selection criteria from \paperI, while the other X-ray source (\dualashorttwo~and \dualbshorttwo) does not pass either of the \LXHAGN~or \hr~criteria.  However, \citet{Liu:2013} find \dualashorttwo~likely to be an AGN based on a one-dimensional analysis of the PSF profiles and X-ray luminosities (soft and hard) that are several times the expected contribution from star-formation, and they find \dualbshorttwo~likely to be an AGN because the soft X-ray luminosity is more than an order of magnitude larger than the expected contribution from star-formation.  Furthermore, slit spectroscopy of these \liudualsysszlet~systems from \citet{Shen:2010c} suggest that two AGN may be present in each system.  While \dualashort~is also in our \full~offset AGN sample, it is rejected from the (unbiased) offset AGN sample and therefore only appears in the dual AGN sample for our analyses.

There is \comerforddualsysszlet~system from \citet{Comerford:2015} that is classified as a dual AGN: \dualc~(\dualcshort).  As with the dual AGN from \citet{Liu:2013}, both of the X-ray AGN in \dualcshort~satisfy the spatially offset criteria from \paperI~($\geq3\sigma$ and $\leq20$ kpc).  The brightest of the two X-ray AGN (\dualcshortone) passes the \LXHAGN~criterion and the weaker source (\dualcshorttwo) is consistent with the \LXHAGN~threshold when accounting for the uncertainty.  These X-ray AGN detections are also consistent with the orientation and separation of AGN photoionized double [OIII]$\lambda$5007 components from slit spectroscopy presented in \citet{Comerford:2012}.

For completeness, we searched within the SDSS AGN-CSC cross-matched sample for additional dual AGN with separations of $\geq3\sigma$ and $\leq20$ kpc, and with velocity separations of less than 600 km s$^{-1}$ (\citealp[e.g. as in ][]{Liu:2012}), finding none.  Finally, we have omitted all sources from the comparison sample that are not within the range of redshifts ($0.025<$~\z~$<0.194$), bolometric luminosities, \lbol~($5.05\times 10^{43}<$~\lbol~$<1.42\times 10^{46}$ erg s$^{-1}$) and projected physical separations, \deltas~($0.04<$~\deltas~$<19.37$ kpc) of the parent sample of \paperI.  This leaves a sample of \parentsdsssz~optically-detected parent AGN and \dualsdsssz~optically-detected AGN in \dualsdsssyssz~dual AGN systems.  We refer to this sample as the \sdss~dual AGN.

\emph{Two caveats about the \sdss~dual AGN sample:} First, the \sdss~dual AGN were originally selected for follow-up imaging ($Chandra+HST$) as dual AGN candidates because of explicit double-peaks in the narrow AGN emission lines of the SDSS spectra and therefore do not represent an unbiased sample of AGN.  Second, the \liudualsysszlet~dual AGN systems from \citet{Liu:2013} would only be selected as offset AGN by the criteria from \paperI, whereas we have adopted their dual AGN interpretations; in this sense, we can not claim the same X-ray properties for the dual AGN as our offset AGN sample.  This choice was made to increase the sample size from one to \dualsdsssysszlet~systems.

\begin{figure*}[t!] $ 
\begin{array}{cc}
\hspace*{-0.1in} \includegraphics[width=3.4in]{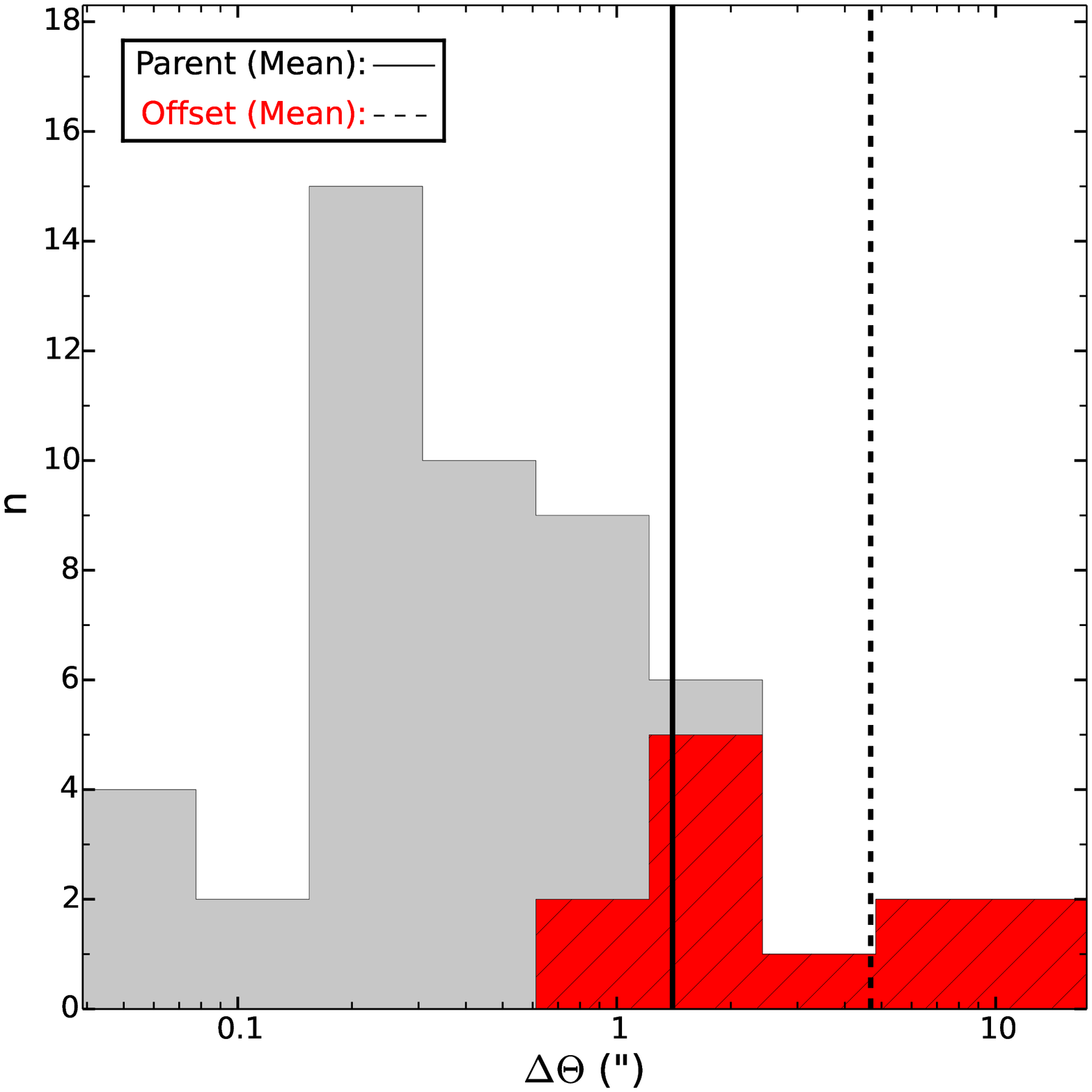} &
\hspace*{0.1in} \includegraphics[width=3.4in]{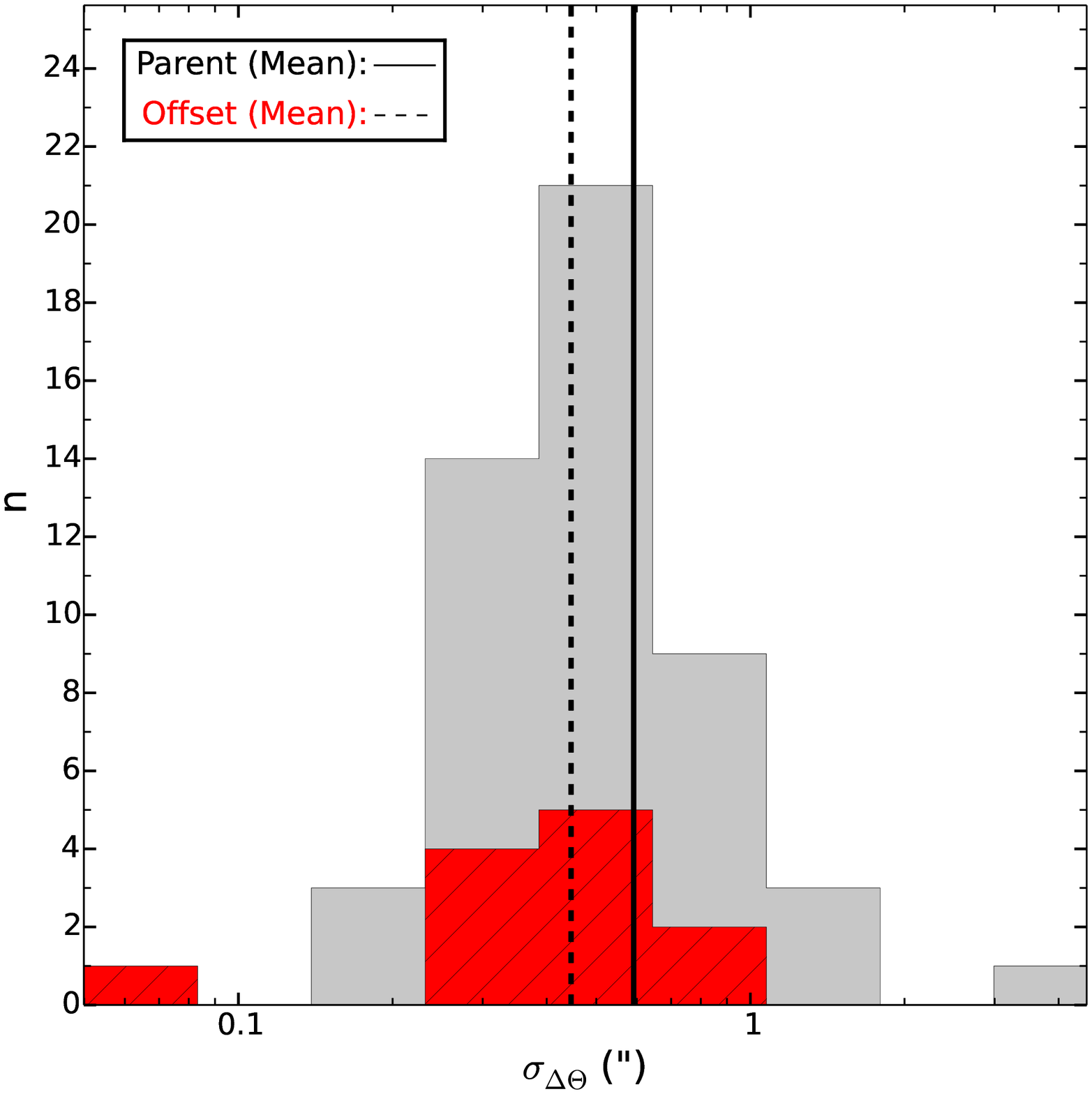}
\vspace*{-0.in}
\end{array} $
\caption{\footnotesize{Histograms of angular offset (left) and angular offset uncertainty (right) for the parent AGN sample (grey) and the offset AGN sample (red, hatched).  The mean values are denoted by vertical solid (parent AGN) and dashed (offset AGN) lines.  Note that the offset AGN sample is distributed toward larger angular separations and smaller uncertainties compared to the parent sample.}
}
\label{fig:hist_bias}
\end{figure*}

\subsubsection{The X-ray-Selected Dual AGN Sample}
\label{sec:bat_dual_agn}

The parent sample of the X-ray-detected dual AGN consists of Burst Alert Telescope (BAT) detections from the 58 Month Survey \citep{Baumgartner:2010} that are crossmatched with AGN \citep{Tueller:2010,Koss:2011}.  The X-ray-selected dual AGN systems we use are the subset of this sample that appear in the \citet{Koss:2012} sample of dual AGN (based on the presence of galaxy companions also hosting an AGN) and with separations of $\leq20$ kpc.  As with the optically-selected dual AGN, the spatial centroids of the AGN host galaxies are separated by $\geq3\sigma$.  

Since the \citet{Koss:2012} sample contains the full sample of BAT AGN in dual AGN systems, we do not search for additional dual AGN within the parent sample.  Furthermore, all of the BAT AGN are within the redshift range of the parent sample.  This yields a parent sample of \parentbatsz~BAT-detected AGN and \dualbatsz~BAT-detected AGN in \dualbatsyssz~dual AGN systems (NGC 6240, Mrk 739, Mrk 463, IRAS 05589+2828, ESO 509-IG 066, IRAS 03219+4031, NGC 3227 and NGC 835).  We refer to this sample as the \bat~dual AGN.  We note that the AGN classifications in the \bat~sample are not uniform and come from a variety of evidence including both optical and X-ray detections.

\section{Analysis}

In this section, we use our offset AGN sample and dual AGN samples to constrain the conditions that affect AGN triggering in galaxy mergers.  Specifically, we examine the AGN merger fractions (Section \ref{sec:frac}), group environments (Section \ref{sec:environ_frac}) and level of nuclear obscuration (Section \ref{sec:LOIII_LX}).

\label{sec:analysis}

\subsection{AGN Merger Fraction}
\label{sec:frac}

In this section, we first derive corrections for the known selection biases in the offset AGN sample (Section \ref{sec:bias}).  Then we investigate the AGN merger fractions as a function of AGN bolometric luminosity, \lbol~(Section \ref{sec:lbol_frac}) and projected nuclear physical separation, \deltas~(Section \ref{sec:phys_sep_frac}).  In each case, we do so for both the offset AGN fractions (\offsetfrac) and the dual AGN fractions (\dualfracsdss~and \dualfracbat).  To reduce the statistical uncertainty, fractions are only shown in each bin if the number of parent AGN is at least \binmin~(which corresponds to $>1\sigma$ confidence in the Poisson count statistics but also allows for an adequate number of bins and dynamic range for our analysis).  The binomial distribution, defined by the size of the parent sample and the success rate of offset AGN occurrences within the parent sample, is used to compute the lower and upper quantiles defining the $68.27\%$ confidence interval around all fraction values.  Uncertainties associated with the functional parameterizations of fractions are the $68.27\%$ quantiles surrounding the median value of each parameter distribution obtained by adding simulated random uncertainties (also drawn from the binomial distribution) and refitting until the uncertainties converge.

\subsubsection{Correcting for Selection Biases}
\label{sec:bias}

The sample of offset AGN was uniformly defined by requiring that the angular offset between the AGN and galaxy core or secondary AGN, \deltatheta, be three or more times its standard uncertainty, \deltathetasig: \deltatheta$\ge3\times$\deltathetasig~(see \paperI~for details).  Therefore, from the parent AGN sample, selection of those with real spatial offsets depends directly on only the two parameters \deltatheta~and \deltathetasig.  Figure \ref{fig:hist_bias} shows the results of our selection process on the distributions of \deltatheta~and \deltathetasig~for both the \infib~and \outfib~subsamples, where the offset AGN have a mean \deltatheta~value greater than that of the parent sample, and a mean \deltathetasig~value less than that of the parent sample.  We have used a two-sample Kolmogorov-Smirnov (KS) test to determine the null hypothesis probability that the parent and offset AGN sample values come from the same distribution (\pnull).  For \deltatheta, the small value of \pnull~$=0.014\%$ strongly suggests that the offset AGN are biased toward large values of \deltatheta~and that we have likely missed offset AGN with small values of \deltatheta\footnotemark[4].  Therefore, the selection has effectively introduced a minimum \deltatheta~limit in the offset AGN sample.  For \deltathetasig, on the other hand, the value of \pnull~$=96.12\%$ suggests that the selection is relatively insensitive to \deltathetasig.  

\footnotetext[4]{We remark that a bias toward large \deltatheta~may introduce a bias toward small redshifts.  However, a two-sample KS-test does not provide statistically significant evidence for the offset AGN to be biased toward small redshifts compared to the parent sample (\pnull~$=76\%$).}

To account for these direct selection effects, we have utilized Monte-Carlo simulations similar to those described in \paperI.  In short, the simulations produce offset nuclei with random projected physical separations ($|$\deltassim$|$~$\leq20$ kpc), redshifts ($0<$~\zsim~$<0.21$) and relative positional uncertainties ($0''<$~\deltathetasigsim~$<0\farcs5$) that are drawn from uniform distributions.  Projecting \deltassim~onto the sky based on random orientations and phases, and then scaling to \zsim~yields a simulated angular offset, \deltathetasim.  Combined with \deltathetasigsim, we selected offset nuclei using the same procedure as in \paperI.  

We then calculated the recovered fraction of simulated offset nuclei as a function of \deltathetasim~and \deltathetasigsim.  Values of \offsetfrac~have been corrected for biases introduced by large values of \deltatheta~and small values of \deltathetasig~based on the results of these simulations.  Specifically, in each bin of \lbol~or \deltas, we identified a simulated offset nucleus with values of \deltathetasim~and \deltathetasigsim~that most closely match \deltatheta~and \deltathetasig, respectively.  Then, we divided the observed AGN merger fraction by the average recovery fraction of the matched simulated nuclei.  The ranges of correction factors are quoted in Sections \ref{sec:lbol_frac} and \ref{sec:phys_sep_frac}.  The simulated recovery fractions should not be taken as estimates of the absolute recovery fractions because the true distributions of the parent sample parameters are unlikely to be uniform as in our simulations.  Therefore, they only provide estimates of the relative fractions as a function of \deltathetasim~and \deltathetasigsim.  Since the corrections are relative, we have normalized them to unity (i.e. no correction) at the largest values of \deltathetasim~and \deltathetasigsim.  We also caution that the corrections based on these simulations do not account for any potential indirect biases which are instead discussed individually in each subsection as footnotes.  We do not correct the values of \dualfracsdss~and \dualfracbat~since the dual AGN positional measurements and uncertainties are not uniformly measured (Section \ref{sec:dual_agn}). 

\subsubsection{Dependence of the AGN Merger Fraction on AGN Bolometric Luminosity}
\label{sec:lbol_frac}

Numerical simulations of galaxy mergers have predicted that the dependence on mergers for AGN triggering is positively correlated with the AGN bolometric luminosity \citep{Hopkins:2009c,Steinborn:2016}, and observational evidence of morphological disturbances in the host galaxies of high luminosity AGN supports these predictions \citep{Schawinski:2012,Treister:2012,Glikman:2015}.  However, other studies that have examined the morphological traits of AGN and non-AGN host galaxies find no statistical differences between samples at low and high AGN luminosities \citep{Georgakakis:2009,Kocevski:2012,Simmons:2012,Villforth:2014,Mechtley:2015,Villforth:2016}.  These null-results imply that AGN triggering is relatively independent of galaxy mergers such that internal instabilities play a comparable if not stronger role in SMBH growth.  While these studies used different procedures and tests, their results are qualitatively in disagreement about the role that galaxy mergers play in triggering AGN as a function of AGN luminosity.  Therefore, we use our systematically selected sample of galaxy mergers to address the role that AGN bolometric luminosity has on the AGN merger fraction.

\begin{figure}$ 
\begin{array}{c}
\vspace*{-0.48in} \hspace*{-0.2in} \includegraphics[width=0.5\textwidth]{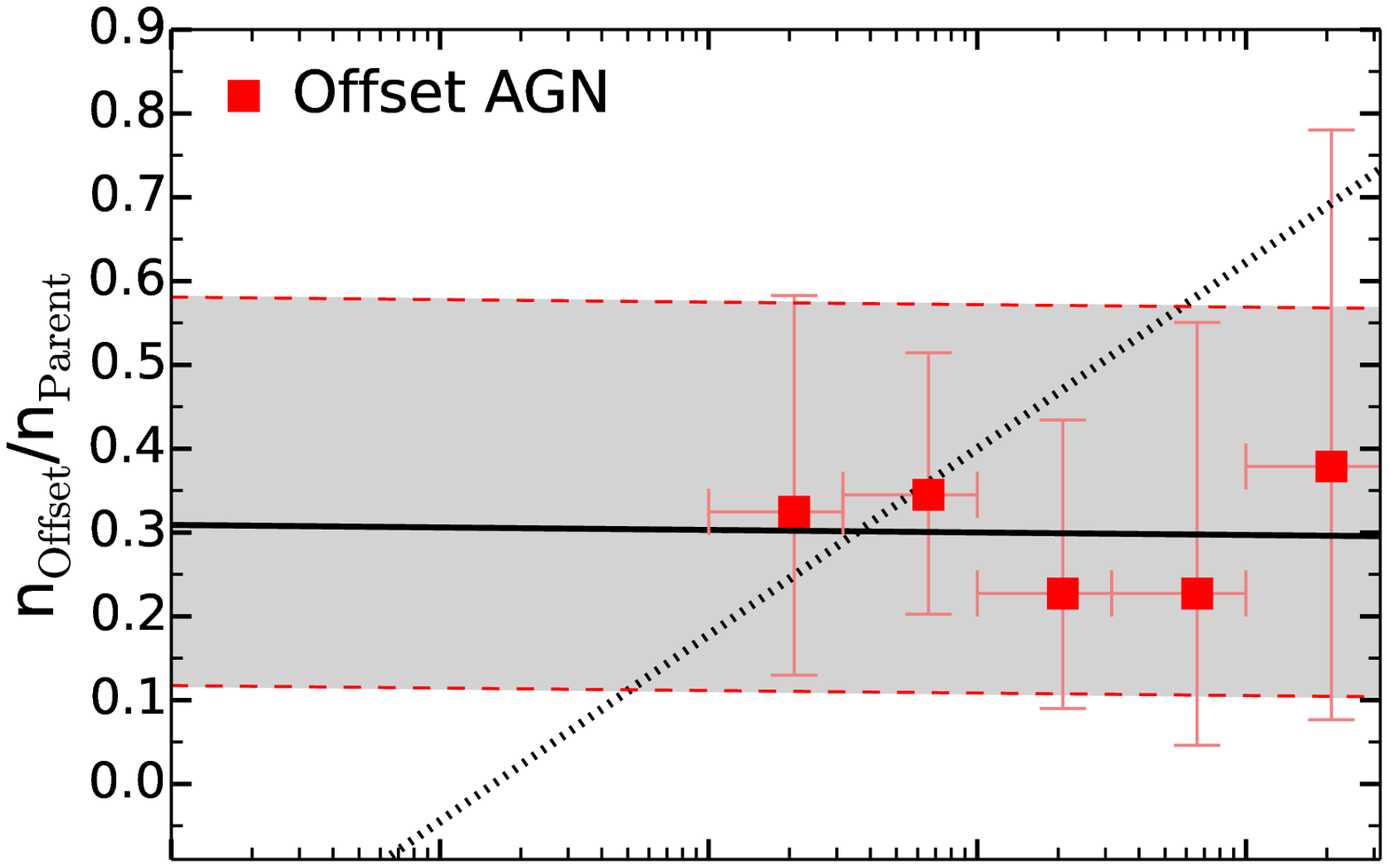} \\
\hspace*{-0.2in} \includegraphics[width=0.5\textwidth]{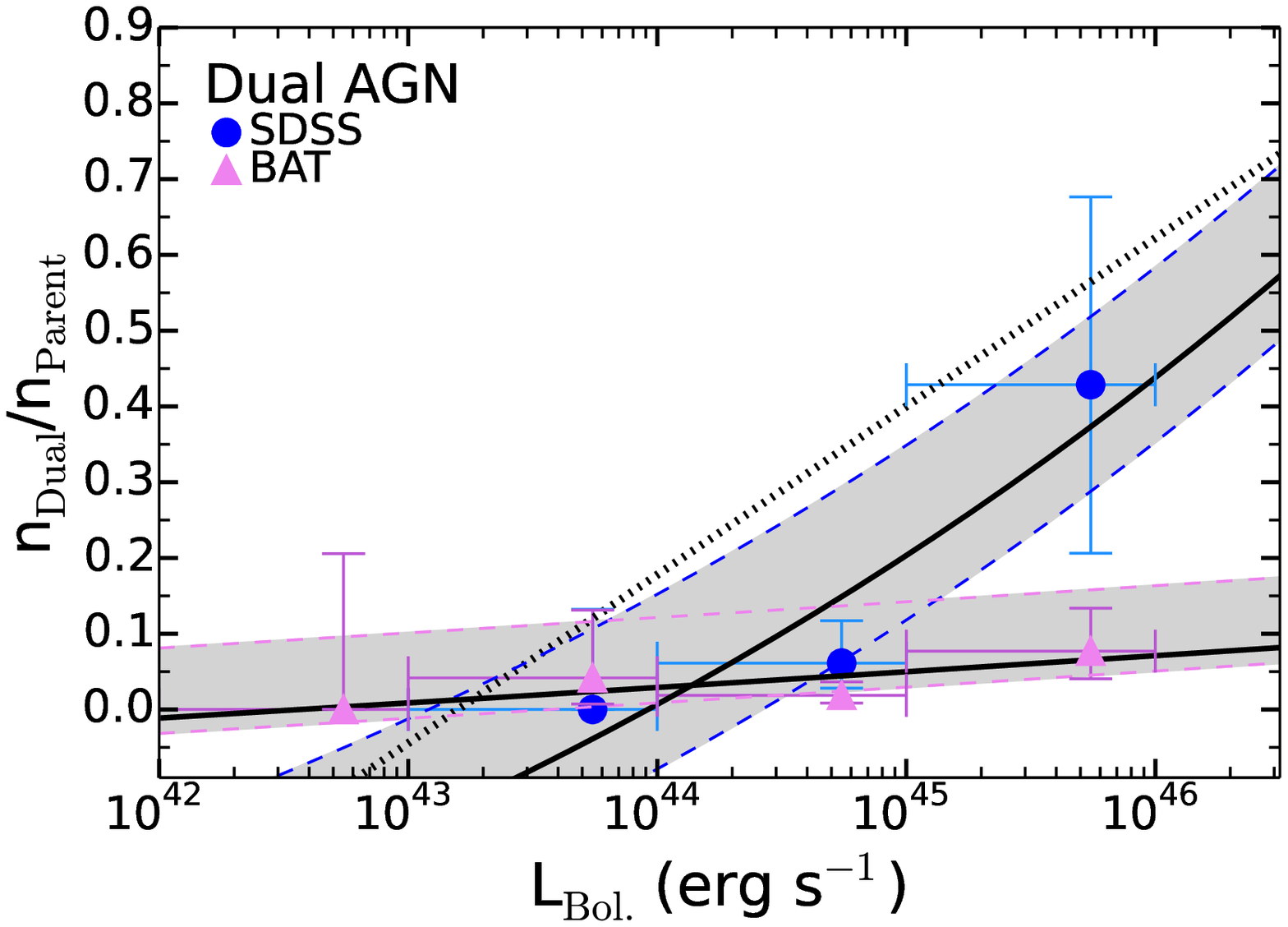}
\vspace*{-0.1in}
\end{array} $
\caption{\footnotesize{Top: the number of bias-corrected offset AGN (red squares) in a given \lbol~bin out of the number of parent AGN in that same bin.  Bottom: same as the top panel but for the SDSS (blue circles) and BAT (violet triangles) dual AGN.  Vertical error bars correspond to $1\sigma$ binomial uncertainties, and horizontal error bars denote the bin width.  The best-fit power-law functions are denoted by the black, solid lines, with the upper and lower $1\sigma$ confidence regimes represented by the grey-shading and bounded by the dashed lines of corresponding color (see Section \ref{sec:frac} for a description of uncertainty estimates).  The dotted, black line represents the linear fit to the fraction of AGN in mergers from \citet{Treister:2012}.  Note that the offset AGN fraction shows no evolution with bolometric luminosity while the dual AGN fractions do show an evolution at $>1\sigma$ significance, and the \sdss~dual AGN fractions most closely match the positive result from \citet{Treister:2012}.}
}
\label{fig:frac_Lbol_OIII}
\end{figure}

Our estimates of \lbol~for the parent sample of the offset AGN and the parent sample of the \sdss~dual AGN are calculated from the extinction-corrected \oiii~luminosity (\LOIII; from \citealp{Oh:2011} based on Balmer decrements) since the original AGN identifications of both samples were based on optical emission lines and are not affected by significant obscuration due to dust (see Section \ref{sec:LOIII_LX})\footnotemark[5].  We use a bolometric correction of \lbol~$=3500$\LOIII~\citep{Heckman:2004}.  Since the companions of the \bat~dual AGN were identified from non-uniform selection criteria, ultra-hard ($14-195$ keV) X-ray luminosities (\LXU) are available for only one AGN in each system.  Therefore, we have chosen to derive \lbol~from \LXU~using the bolometric correction of \lbol~$=15$\LXU~\citep{Vasudevan:2009}.  While this choice means we can only use eight of the \dualbatsz~AGN in the \bat~systems, the use of ultra-hard X-rays mitigates the effect of nuclear obscuration \citep{Koss:2011b}.

\footnotetext[5]{While the use of \LOIII~for deriving \lbol~avoids the effects of serious nuclear obscuration encountered with \LXHAGN, the $1\farcs5$ SDSS fiber radius means that the \oiii~emission is not well-constrained.  This may be a problem in cases where more than one AGN is present and the \oiii~emission originates from a different location that of the X-ray AGN emission.  However, even in this case, our analysis is still tracing AGN bolometric luminosities for systems in which an X-ray AGN is present in a galaxy merger.}

Figure \ref{fig:frac_Lbol_OIII} (top) shows \offsetfrac, calculated as the number of offset AGN (\noff) out of the number of parent AGN (\npar), as a function of \lbol.  Figure \ref{fig:frac_Lbol_OIII} (bottom) shows \dualfracsdss~and \dualfracbat~as a function of \lbol, where each is computed in the same manner as \offsetfrac.  To optimize the combination of signal and binning resolution, we have chosen bin sizes individually for each sample.  Due to the multiple orders of magnitude spanned by \lbol, the bins have been given logarithmically uniform spacing with sizes of 0.5 dex for the offset AGN sample and 1 dex for both dual AGN samples.  After implementing the threshold of $\geq$\binmin~parent AGN in each bin, the resulting \lbol~bin ranges are $10^{44}-10^{46.5}$ erg s$^{-1}$ for \offsetfrac, $10^{43}-10^{46}$ erg s$^{-1}$ for \dualfracsdss~and $10^{42}-10^{46}$ erg s$^{-1}$ for \dualfracbat.  Therefore, to examine both the offset and dual AGN samples over the same \lbol~values, we have chosen the plotting range of \lbol~$=10^{42}-10^{46.5}$ erg s$^{-1}$.

From \lbol~$=10^{44}-10^{46.5}$ erg s$^{-1}$, \offsetfrac~is adequately fit by a power-law function ($y=ax^{b}$) with parameter values of $a=$~\offsetLBolA~and $b=$~\offsetLBolB.  This fit corresponds to a slope with a \offsetLBol~significance from zero.  For comparison, the correction factors (Section \ref{sec:bias}) range from 1.14 to 1.21, and the best-fit power-law parameters for the uncorrected fractions are $a=$~\offsetLBolAnocorr~and $b=$~\offsetLBolBnocorr, corresponding to a slope with a \offsetLBolnocorr~significance from zero.  To test the implications of using the \outfib~subsample (Section \ref{sec:offset_agn}), we remove the \outfib~sources and find that doing so does not change the qualitative result that no significant change in \offsetfrac~is seen as a function of \lbol.  We also find that removing sources with \LXHAGN~$<10^{41}$ erg s$^{-1}$ or that are LINERS also has no effect on this qualitative result.

From \lbol~$=10^{43}-10^{46}$ erg s$^{-1}$, \dualfracsdss~is adequately fit by a power-law function with parameter values of $a=$~\dualsdssLBolA~and $b=$~\dualsdssLBolB.  Compared to \offsetfrac, this fit corresponds to a steeper slope with a \dualsdssLBol~significance from zero\footnotemark[6].  From \lbol~$=10^{42}-10^{46}$ erg s$^{-1}$, \dualfracbat~is also adequately fit by a power-law function with parameter values of $a=$~\dualbatLBolA~and $b=$~\dualbatLBolB.  Compared to \dualfracsdss, this fit corresponds to a shallower slope, though it still has a \dualbatLBol~significance from zero.  While the current data offers a null result for \offsetfrac, a similar increase from low to high \lbol~can not be ruled out due to the significant fit uncertainties.

\footnotetext[6]{A correlation between \lbol~and small values of \deltathetasig~may be expected since \deltathetasig~is determined in part by the AGN flux.  However, we find that the correlation statistic between \lbol~and \deltathetasig~is small for all samples.  Therefore, we argue that a correlation between \lbol~and the offset or dual AGN fractions is not driven by a selection bias.}

For comparison, in both panels of Figure \ref{fig:frac_Lbol_OIII} we show the best-fit linear function to the AGN merger fraction from \citet{Treister:2012}.  The parent sample used in \citet{Treister:2012} consists of AGN identified from X-ray, infrared and spectroscopic surveys, and the galaxy merger systems were taken from samples identified by visual classification.  As a result, the number of AGN in each system (offset or dual AGN) is usually not possible to determine.  In the top panel, we see that the function from \citet{Treister:2012} is consistent with \offsetfrac~below \lbol~$=10^{45}$ erg s$^{-1}$ while over-predicting \offsetfrac~at higher \lbol~by $\sim1\sigma$. This result provides tentative evidence that \offsetfrac~behaves differently from the \citet{Treister:2012} function at high \lbol~values, though the substantial uncertainties prohibit a firm conclusion.  The \offsetfrac~\lbol~values plotted also do not extend down to the luminosities of $\sim10^{42}$ erg s$^{-1}$, and thus the behavior of \offsetfrac~compared to the lower \lbol~end of the \citet{Treister:2012} sample is not known.  The stronger positive correlations seen in the dual AGN (relative to the offset AGN) are in better qualitative agreement with the function from \citet{Treister:2012}, though the \bat-sample is generally over-predicted while the \sdss~sample is in agreement to within the $1\sigma$ uncertainties over nearly the full \lbol~range plotted. 

\subsubsection{Dependence of the AGN Merger Fraction on Projected Physical Separation}
\label{sec:phys_sep_frac}

Numerical simulations have predicted that, in an evolving galaxy merger, the probability of observing an AGN increases with decreasing separation of the two SMBHs from the progenitor galaxies \citep{Van_Wassenhove:2012,Blecha:2013,Stickley:2014}.  Observational evidence of rising AGN merger fractions with decreasing nuclear separation supports these predictions \citep{Iwasawa:2011,Ellison:2011,Silverman:2011,Liu:2012a,Satyapal:2014}.  However, the galaxy mergers from those samples were identified from galaxy pairs, thereby limiting the nuclear separations to larger values (several kpc or greater) so that individual galaxies can be distinguished.  Simulations, on the other hand, predict that the AGN observability continues to increase significantly below 1 kpc, and previous observational studies have not been able to probe the small separation regime where the dependence of the AGN merger fraction on nuclear separation is predicted to peak.  Therefore, we use our sample of offset AGN with resolved X-ray AGN offsets from $20$ to $\sim0.8$ kpc to examine the AGN merger fraction from early to late merger stages.  We have adopted the physical separations presented in \citet{Liu:2013} and \citet{Comerford:2015} for the \sdss~dual AGN and those presented in \citet{Koss:2012} for the \bat~dual AGN.

\begin{figure}[t!]$ 
\begin{array}{c}
\vspace*{-0.46in} \hspace*{-0.2in} \includegraphics[width=0.5\textwidth]{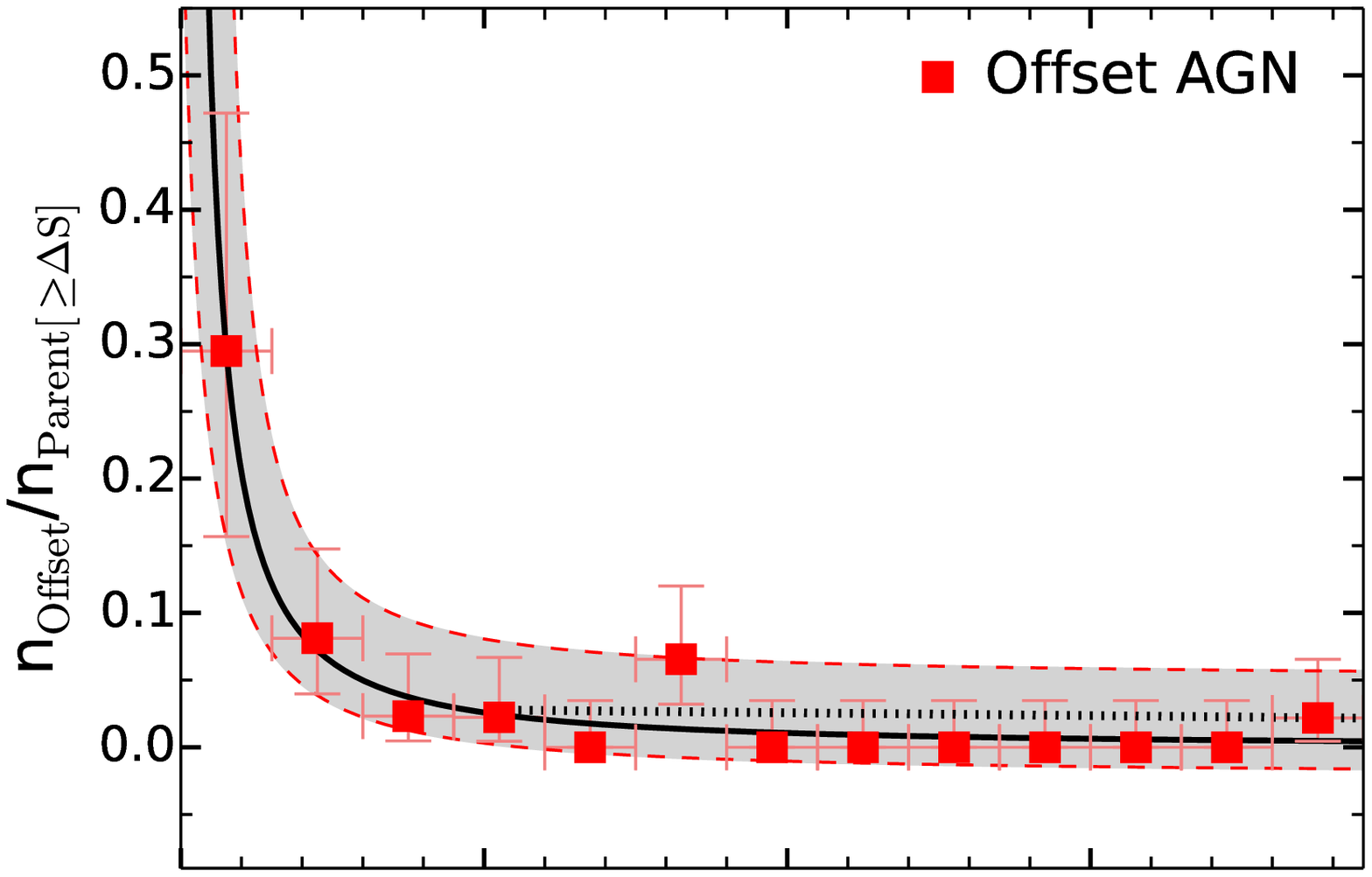} \\
\hspace*{-0.2in} \includegraphics[width=0.5\textwidth]{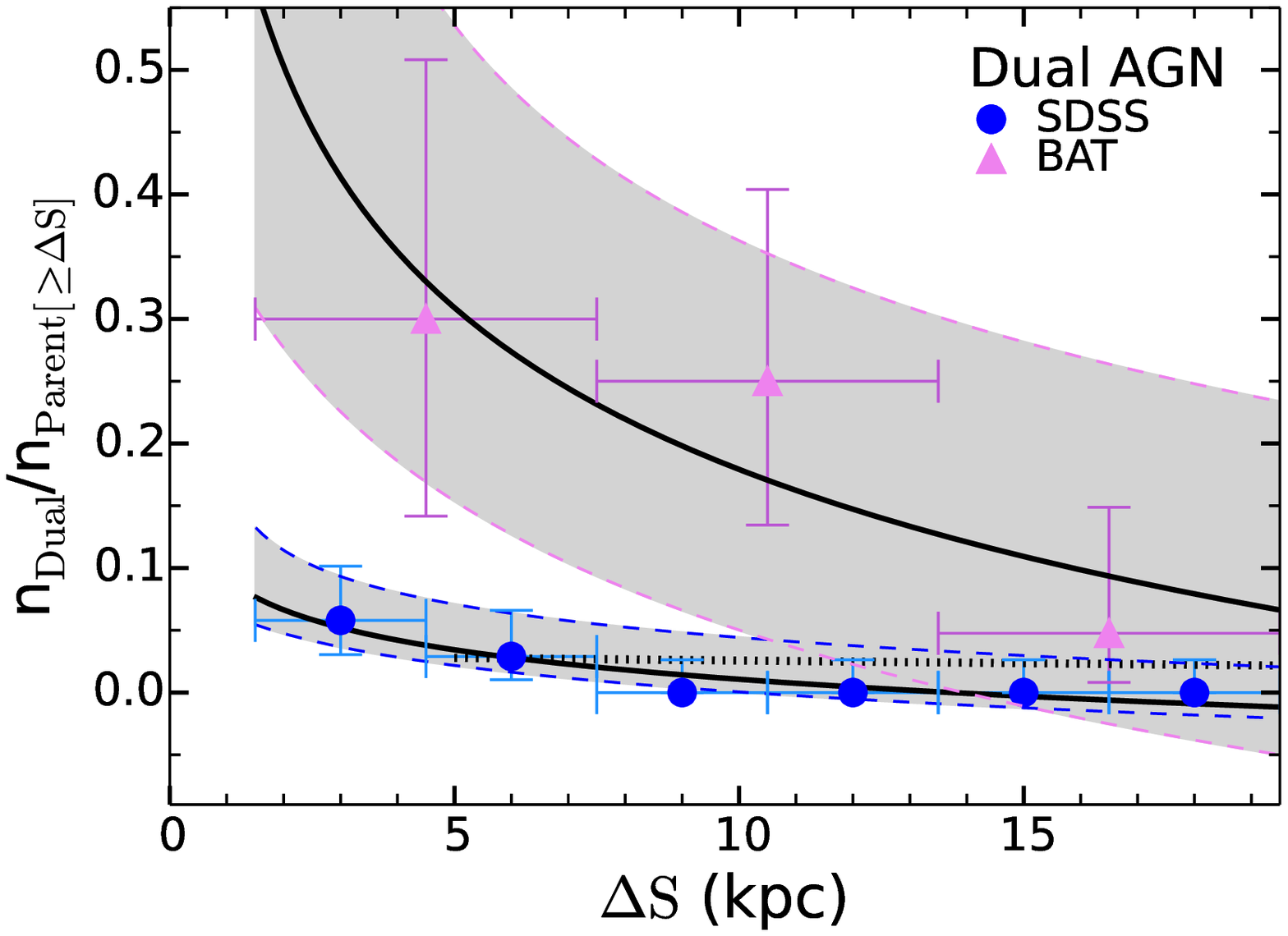}
\vspace*{-0.1in}
\end{array} $
\caption{\footnotesize{Top: the number of bias-corrected offset AGN (red squares) in a given \deltas~bin out of the subset of the parent AGN sample with positional uncertainties small enough to resolve down to the mean separation of that same \deltas~bin at the host galaxy redshift.  Bottom: same as the top panel but for the SDSS (blue circles) and BAT (violet triangles) dual AGN.  Vertical error bars correspond to $1\sigma$ binomial uncertainties, and horizontal errors bars denote the bin width.   The best-fit power-law functions are denoted by the black, solid lines, with the upper and lower $1\sigma$ confidence regimes represented by the grey-shading and bounded by the dashed lines of corresponding color (see Section \ref{sec:frac} for a description of uncertainty estimates).  The functions are only shown down to the lower end of the smallest \deltas~bin.  The dotted, black line represents the \citet{Satyapal:2014} AGN merger fraction trend (see Section \ref{sec:phys_sep_frac} for details).  The line is drawn down to $5$ kpc, representing the approximate minimum resolvable separation of typical galaxy pair samples from the SDSS.  Note that both the offset and dual AGN fractions show an evolution with decreasing nuclear separation at $>1\sigma$ significance and are consistent with the results from \citet{Satyapal:2014} at larger separations.}
}
\label{fig:frac_phys_sep}
\end{figure}

Figure \ref{fig:frac_phys_sep} (top) shows \offsetfrac~as a function of \deltas.  Since values of \deltas~are not available for the parent AGN sample, we have computed the minimum projected physical separation that could potentially be resolvable (by $\geq3\sigma_{\Delta \Theta}$) based on the \deltathetasig~and \z~values for each parent AGN: \deltaspot.  The values of \offsetfrac~in each \deltas~bin are then the number of offset AGN with \deltas~within that bin (\noff) out of the number of parent AGN with \deltaspot~less than or equal to the mean value of that bin (\nparcum).  Note that a parent AGN may be in multiple bins using this approach.  Figure \ref{fig:frac_phys_sep} (bottom) shows \dualfracsdss~and \dualfracbat~as a function of \deltas, where each is computed in the same manner as \offsetfrac.  As in Section \ref{sec:lbol_frac}, we have chosen bin sizes individually for each sample to optimize the combination of signal and binning resolution.  The bins have been given linearly uniform spacing with sizes of 1.5 kpc for the offset AGN sample, 3 kpc for the \sdss~dual AGN sample, and 6 kpc for the \bat~dual AGN sample.  After implementing the threshold of $\geq$\binmin~parent AGN in each bin, the resulting \deltas~bin ranges are $0-19.5$ kpc for \offsetfrac~and $1.5-19.5$ kpc for both \dualfracsdss~and \dualfracbat.  Therefore, to examine both the offset and dual AGN samples over the same \deltas~values, we have chosen the plotting range of \deltas~$=0-19.5$ kpc.

From \deltas~$=0-19.5$ kpc, \offsetfrac~is adequately fit by a power-law function with parameter values of $a=$~\offsetPhySepA~and $b=$~\offsetPhySepB.  This fit corresponds to a slope with \offsetPhySep~significance from zero.  The correction factors (Section \ref{sec:bias}) range from 0 for fractions in bins of large \deltas~to 1.28 in the bin of smallest \deltas.  For comparison, the best-fit power-law parameters for the uncorrected fractions are $a=$~\offsetPhySepAnocorr~and $b=$~\offsetPhySepBnocorr, corresponding to a slope with  \offsetPhySepnocorr~significance from zero.  As in Section \ref{sec:lbol_frac}, we test the implications of including the \outfib~subsample by removing those sources, finding that doing so has no qualitative effect on the significant rise in \offsetfrac~at small physical separations since the \outfib~offset AGN have physical separations larger than the small-separation regime in which the rapid change in \offsetfrac~is seen.  This test also confirms that the correlation is not due to the artifact introduced by the fiber size.  Likewise, removal of the sources with \LXHAGN~$<10^{41}$ erg s$^{-1}$ or those that are LINERS has no qualitative effect on the significant rise in \offsetfrac~at small physical separations.

From \deltas~$=0-19.5$ kpc, \dualfracsdss~is adequately fit by a power-law function with parameter values of $a=$~\dualsdssPhySepA~and $b=$~\dualsdssPhySepB.  Compared to \offsetfrac, this fit corresponds to a shallower slope, but it is still has a \dualsdssPhySep~significance from zero.  From \deltas~$=1.5-19.5$ kpc, \dualfracbat~is also adequately fit by a power-law function with parameter values of $a=$~\dualbatPhySepA~and $b=$~\dualbatPhySepB.  Compared to \dualfracsdss, this fit is steeper (though still shallower than \offsetfrac) and has a \dualbatPhySep~significance from zero.  We note that \offsetfrac~dramatically increases at small separations when compared to \dualfracsdss~and \dualfracbat~(as indicated by their respective magnitudes of $b$).  This difference may indicate that the value of $b$ in the \offsetfrac~trend is strongly driven by the data point at \deltas$<1.5$ kpc which is lacking in the \sdss~and \bat~dual AGN samples.

For comparison, in both panels of Figure \ref{fig:frac_phys_sep} we show a line representing the trend seen in the AGN merger fraction as a function of projected pair separation from \citet{Satyapal:2014} based on the \wise~color cut ($W1-W2>0.8$) that has been empirically shown to select infrared bright AGN with high reliability.  In particular, the line shows the slope made by the change in the AGN merger fraction in their pre-coalescence sample from the largest separation data point to the smallest separation data point plotted in their Figure 2.  The \citet{Satyapal:2014} sample utilizes galaxy mergers selected as galaxy pairs (not distinguishing between the offset and dual AGN scenarios) that extends from a minimum separation of $\sim5$ kpc (based on the redshift range and the SDSS fiber size) out to \deltas~$=80$ kpc \citep{Ellison:2013}.  Above \deltas$\approx12$ kpc, \offsetfrac, \dualfracsdss~and \dualfracbat~are all consistent with the \citet{Satyapal:2014} sample, while all three merger fractions would be under-predicted by a linear extrapolation of this slope at \deltas$<3$ kpc.

\subsection{Environments of AGN hosts in Mergers}
\label{sec:environ_frac}

Observationally, galaxies that exist in dense environments, such as groups or clusters, generally have redder colors (due to suppressed star-formation) and more elliptical morphologies compared to galaxies in less dense environments \citep{Dressler:1980,Kauffmann:2004,Trinh:2013}.  One route for dense environments to drive galaxies toward redder colors and elliptical morphologies is through interactions or mergers between galaxies.  While the probability of a direct merger is low among satellite galaxies in clusters due to their high relative velocities, galaxy groups are the environments most likely to contain merging systems \citep{McIntosh:2008}.  Therefore, we may expect to find a higher fraction of offset and dual AGN in group environments.  

We use our sample of offset and dual AGN to test this prediction by quantifying the density of their environments.  We do so by first taking a catalogue of galaxy group members described in \citet{Wetzel:2012}.  The catalogue was constructed by applying host dark matter halo and satellite dark matter halo mass functions \citep{Tinker:2008,Tinker:Wetzel:2010} to galaxies from the SDSS DR7.  Based on the halo mass functions of each galaxy, the central most massive galaxy of a group and its satellites can be identified.  Thus, every galaxy in the catalogue has an assigned number of group members, \ngroup, with masses above $5\times 10^{9}$ M$_{\odot}$ (below this mass limit the method is not sensitive) which we take as a parameterization of the group density.  We then matched the offset AGN parent sample and the dual AGN parent samples with the catalogue to assign them \ngroup~values.  Since the redshift range of the galaxy group catalogue is limited to $z<0.1$, only a subset of the AGN in our samples will have \ngroup~values.

\begin{figure}
\hspace*{-0.2in} \includegraphics[width=3.45in]{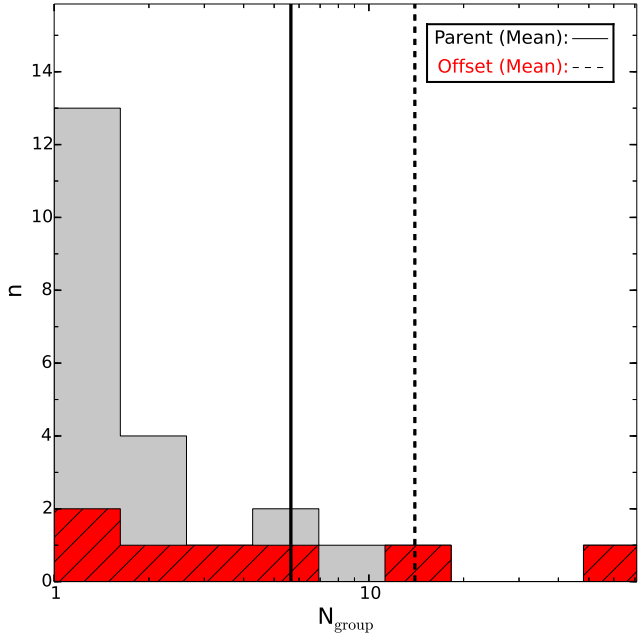}
\vspace*{-0.in}
\caption{\footnotesize{The distribution of \ngroup~values for the parent AGN (grey) and the offset AGN (red, hatched).  The mean values are denoted by vertical solid (parent AGN) and dashed (offset AGN) lines.  Note that the two samples show no significant evidence for being drawn from different distributions.}
}
\label{fig:hist_ngroup}
\end{figure}

Figure \ref{fig:hist_ngroup} compares the \ngroup~distributions for the parent and offset AGN.  While the mean value of the offset AGN (14.0) is larger than that of the parent AGN (5.6), the difference is not at a significant level based on a two-sample KS test (\pnull~$=59\%$).  As in Sections \ref{sec:lbol_frac} and \ref{sec:phys_sep_frac}, we test the implications of including the \outfib~subsample by removing those sources, finding that doing so does not change the qualitative result that no significant  difference is seen between the two samples.  Likewise, removal of the sources with \LXHAGN~$<10^{41}$ erg s$^{-1}$ or those that are LINERS has no qualitative effect on this result.  Too few of the \sdss~and \bat~AGN are assigned \ngroup~values to statistically examine their environments.

While the number statistics are small, this result suggests that offset AGN tend to be in environments similar to the general Type 2 AGN population, or at least that the difference is small enough to be undetectable in our sample.  This suggestion implies that the types of mergers leading to the spatially offset nature of an X-ray AGN within the parent sample are not linked with over-dense environments compared to the parent sample.  In fact, the environments of both the parent and offset AGN have relatively small densities of group members compared to cluster environments (\ngroup~$>50$).  However, these densities only correspond to group members with masses above $5\times 10^{9}$ M$_{\odot}$, suggesting that mergers with lower-mass galaxies may play a role in producing the offset AGN systems.  The implications of this merger mass ratio effect are discussed in Section \ref{sec:disc_single_dual}.

\subsection{Optical Versus X-ray Luminosities}
\label{sec:LOIII_LX}

The nuclear regions of merging galaxy systems can potentially be heavily obscured as gas and dust will naturally settle toward the regions of largest gravitational potential.  Since the \oiii~emission line originates far enough from the SMBH to not be subject to nuclear obscuration, comparison with the X-ray luminosity, which does originate near the SMBH accretion disk, can potentially reveal the presence of nuclear enshrouding material.  

\begin{figure}[t!] $ 
\begin{array}{c}
\hspace*{-0.1in} \includegraphics[width=3.45in]{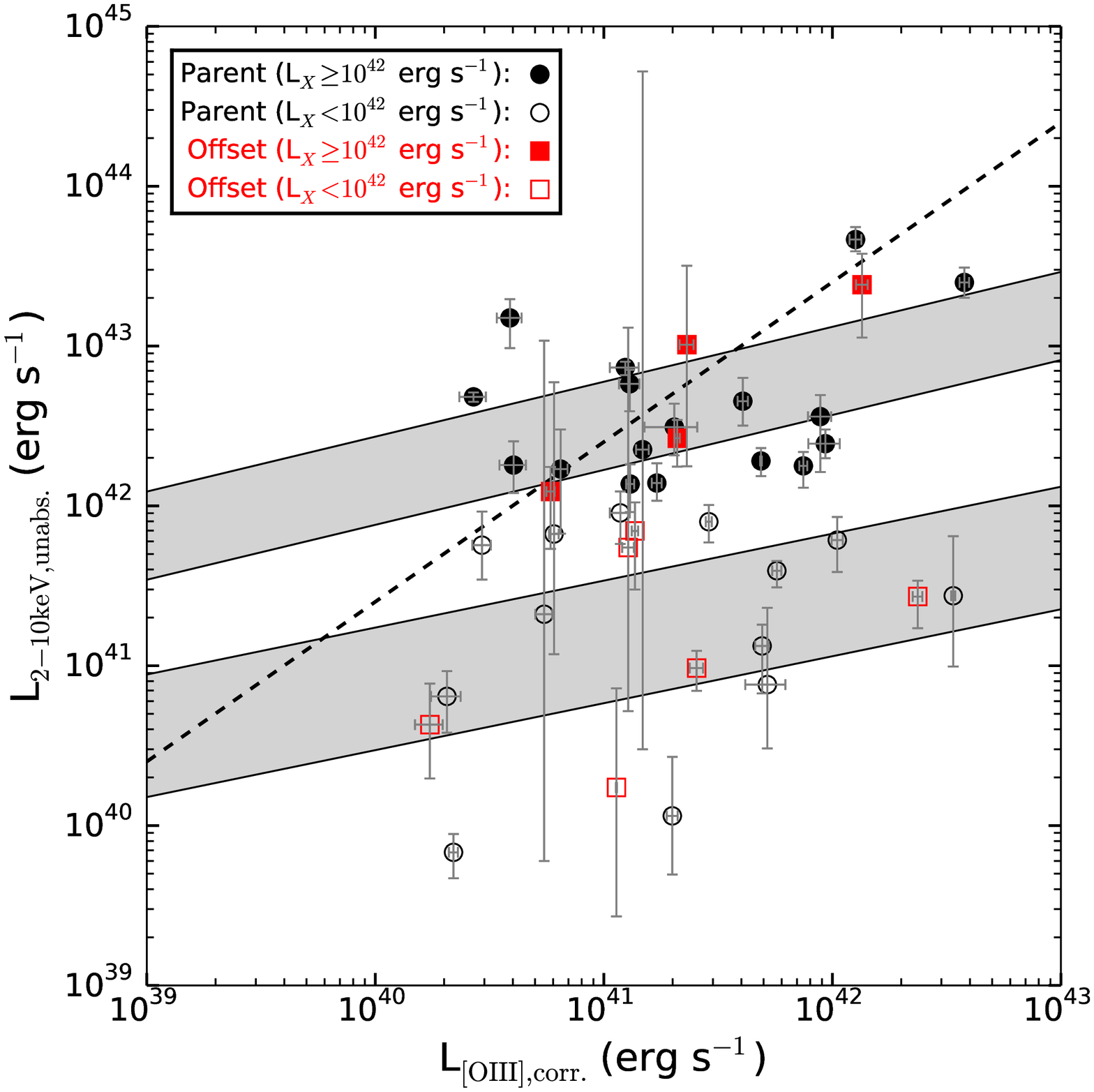}
\end{array} $
\caption{\footnotesize{Unabsorbed 2-10 keV luminosity plotted against extinction-corrected \oiii~luminosity.  The black circles represent the parent AGN sample and the red squares represent the offset AGN sample.  The filled symbols denote the subsample that passes the \LXHAGN~threshold while the open symbols denote the subsample that only passes the \hr~threshold.  The upper and lower standard deviation bounds around the best-fitting linear functions are shown as grey-shaded regions for the \LXHAGN-selected parent AGN (top) and for the \hr-selected parent AGN (bottom).  The mean value for the merger-independent sample of \citet{Trichas:2012} is shown with the black, dashed line.  Note that the \hr-selected subsample has systematically lower values of \LXHAGN~compared to the \LXHAGN-selected subsample.
}
}
\label{fig:plot_LX_LOIII}
\end{figure}

Therefore, we have tested for this effect in Figure \ref{fig:plot_LX_LOIII} by plotting \LXHAGN~against \LOIII~for the parent AGN sample and offset AGN sample.  We have also shown in Figure \ref{fig:plot_LX_LOIII} the mean value of a Type 2 AGN sample selected independently of mergers \citep{Trichas:2012,Liu:2013}.  The effect of our \LXHAGN~and hardness ratio (\hr) selection criteria can be seen in Figure \ref{fig:plot_LX_LOIII}.  The \LXHAGN-selected subsample has \LXHAGN/\LOIII~ratios similar to the merger-independent sample, whereas the \hr-selected subsample has systematically lower values of \LXHAGN~for a given value of \LOIII.  Since the \LXHAGN-selected subsample has an average column density that is larger by $1.14$ dex than that of the \hr-selected subsample, this difference between the two subsamples is consistent with underestimates of the column densities in the latter (assuming that the distributions of intrinsic X-ray spectral slopes and intrinsic absorbing columns are the same).  Indeed, the relatively simple absorbed power-law fit to the Chandra spectra (\paperI) may be under-estimating the column densities in these sources due to low counts.  

While the mean \LXHAGN/\LOIII~ratios of the offset AGN are lower than those of the parent AGN for the \LXHAGN-selected subsample ($23.95$ versus $45.36$) and the \hr-selected subsample ($2.08$ versus $3.28$), they are in agreement to within the scatter.  Furthermore, the two-sample KS test results of \pnull~$=59.9\%$ and \pnull~$=99.9\%$ for the \LXHAGN-selected and \hr-selected subsamples, respectively, do not suggest a significant difference.  Therefore, for AGN with a given value of \LXHAGN~and \LOIII, selection of X-ray sources that are spatially offset from the host galaxy core or a nearby galaxy core does not appear to introduce a bias toward obscuration.  Instead, this result indicates that our selection of offset AGN, from the parent AGN sample, coincides with a tendency to select relatively unobscured systems.  The potential physical implications of this result are discussed in Section \ref{sec:obscuration}.

\section{Discussion}
\label{sec:discussion} 

Several studies have placed estimates on the fraction of AGN hosted by galaxies in mergers or merger-remnants  \citep{Villforth:2014,Georgakakis:2009,Schawinski:2011,Schawinski:2012,Cisternas:2011,Kocevski:2012,Simmons:2012,Ellison:2011,Silverman:2011,Bessiere:2012,Treister:2012}.  However, these studies have not estimated the fractions for the specific scenarios of offset AGN or dual AGN because the methods by which those studies selected galaxy mergers and cross-matched with AGN does not uniformly allow for a distinction between systems in which one or both galaxies hosts an AGN.  The distinction between the two scenarios is a crucial step toward understanding the physics that govern accretion onto SMBHs within galaxy mergers.  Since our selection method requires the AGN to be spatially isolated within the merger, we can measure the number of AGN and constrain the conditions of offset versus dual AGN formation.  In this section, we discuss the offset and dual AGN scenarios in the context of AGN luminosity (Section \ref{sec:disc_single_dual}), merger stage (Section \ref{sec:disc_sep}), the combined effect of AGN luminosity and merger stage (Section \ref{sec:sep_lum}), and finally nuclear obscuration (Section \ref{sec:obscuration}).

\subsection{Triggering of High Luminosity AGN in Mergers}
\label{sec:disc_single_dual}

As mentioned in Section \ref{sec:lbol_frac}, some observational evidence suggests that high-luminosity AGN are preferentially found in galaxies that are interacting, merging, or have merged in the past, while other studies find no evidence for AGN hosts to have different morphologies from inactive galaxies.  In Section \ref{sec:lbol_frac} and Figure \ref{fig:frac_Lbol_OIII}, we used our AGN merger samples (selected independent of galaxy or merger morphology) to address this discrepancy, producing the following results: 

Based on their parameterizations (Section \ref{sec:lbol_frac}), the offset and dual AGN fractions behave differently as a function of \lbol.  The offset AGN fraction displays no significant evolution over the AGN bolometric luminosity range probed (\offsetLBol~significance) as shown in Figure \ref{fig:frac_Lbol_OIII} (top).  Comparatively, the dual AGN fractions show stronger evidence for a statistically significant increase at high AGN bolometric luminosities (\dualbatLBol$-$\dualsdssLBol~significance) as shown in Figure \ref{fig:frac_Lbol_OIII} (bottom).  

These results can be broadly interpreted as in agreement with claims of a positive correlation between the AGN merger fraction and bolometric luminosity.  Specifically, observations of large-scale (10-100 kpc) galaxy pairs have shown that the AGN merger fraction is highest for major mergers \citep{Woods:2007,Ellison:2011}, implying that major mergers are more efficient at removing angular momentum from gas and dust in their host galaxies.  Additionally, \citet{Koss:2012} find that their BAT-selected dual AGN are preferentially found in major mergers.  Therefore, dual AGN host systems may show a preference for major mergers so that enough fuel is available to power both AGN.  For example, of the \dualsdsssysszlet~\sdss~dual AGN systems, two are hosted by major mergers \citep{Shangguan:2016}.  Since the sample from \citet{Treister:2012} extends to high redshifts and was selected based on morphology, that sample may contain a relatively higher fraction of major mergers that allow for easier visual classification, possibly accounting for the agreement between that sample and the comparison SDSS dual AGN sample.  

By contrast, offset AGN may show a preference to reside in minor mergers instead of major mergers.  For example, in Section \ref{sec:environ_frac} we examined the distribution of the number of group members for both the parent and offset AGN, finding no significant evidence for a difference between the two samples (Figure \ref{fig:hist_ngroup}) and that both samples reside in relatively low-density environments.  However, since the procedure used for measuring group members is not sensitive to galaxies with masses $<5\times 10^{9}$ M$ _{\odot}$ \citep{Wetzel:2012}, we are likely missing many lower-mass group members.  Therefore, compared to galaxy mergers seen in dense environments of more massive galaxies, the offset AGN are more likely to be undergoing mergers with galaxies of lower masses, corresponding to mass ratios that fall in the minor merger regime. 

However, we caution that the offset AGN slope is consistent within $1\sigma$ with those of both dual AGN samples (Figure \ref{fig:frac_Lbol_OIII}) so that the true difference between offset and dual AGN evolution with AGN bolometric luminosity is poorly constrained.  We also note that the effect of merger mass ratio on AGN triggering is still poorly understood at small separations, particularly for offset AGN, and may also depend on whether or not loss of angular momentum happens more efficiently in the major or minor galactic stellar core.  Currently, theoretical work has provided ambiguous results, with one model suggesting that the more luminous AGN likely resides in the more massive stellar bulge \citep{Yu:2011}, while a recent simulation of galaxy mergers has suggested that the accretion rate is higher for the AGN in the less massive galaxy \citep{Capelo:2015}.  In Paper III, follow-up imaging of our offset AGN sample with $HST$ will put constraints on these predictions by allowing estimates of the merger mass ratios and SMBH accretion rates.

\subsection{Triggering AGN at Small Nuclear Separations}
\label{sec:disc_sep}

As mentioned in Section \ref{sec:phys_sep_frac}, numerical work predicts that the probability of AGN triggering becomes strongest at separations below $1$ kpc as the two SMBHs dynamically evolve toward the region of greatest gravitational potential along with a significant amount of gas and dust for accretion.  In Section \ref{sec:phys_sep_frac} and Figure \ref{fig:frac_phys_sep}, we used our AGN merger samples (with measured physical separations reaching below $1$ kpc) to test this prediction, producing the following results: 

Evidence for a negative correlation with nuclear separations below $20$ kpc is seen at $>1\sigma$ significance for both the offset AGN fractions (\offsetPhySep) and the dual AGN fractions (\dualbatPhySep$-$\dualsdssPhySep) as shown in Figure \ref{fig:frac_phys_sep}.  While the physical separations of the dual AGN sample do not allow us to see this trend continue below $1.5$ kpc, we are able to do so for the offset AGN sample due to our astrometric registration procedure.  Indeed, the slope magnitude for the offset AGN fractions is larger (by $>2\sigma$) than those of the dual AGN fractions, an indication that the merger fraction rises most strongly at the smallest separations.  A similar, or larger, increase may be seen in the dual AGN sample with a larger sample size or with data points at separations $<1.5$ kpc.  Our finding that the AGN merger fraction rises fastest and peaks (down to our resolution limits) below 1 kpc is consistent with numerical predictions.

At large separations, the offset AGN fractions and the dual AGN fractions are consistent with the slope of the AGN merger sample presented in \citet{Satyapal:2014}.  In fact, both the offset AGN and \sdss~dual AGN fractions are consistent with this slope over the full range shown ($5-20$ kpc).  At large separations, our results are also consistent with other studies \citep{Silverman:2011,Ellison:2011,Liu:2012a}.

When examining the absolute values of the AGN merger fractions, we see that the offset AGN and \sdss~dual AGN fractions (below $5$ kpc) of $13.7^{+6.6}_{-4.9}\%$ and $12.1^{+8.5}_{-5.7}\%$, respectively, are in agreement when accounting for their $1\sigma$ uncertainties.  That the offset AGN and \sdss~dual AGN fractions are consistent at small separations may be indicating that, among similarly selected samples, dual AGN triggering becomes more likely at smaller separations and comparable to single AGN triggering.  This result is consistent with the expectation that there is an increased supply of gas for accretion at smaller nuclear separations.  The generally larger values of the \bat~dual fraction may indicate a stronger overall dependence of AGN triggering on nuclear separation in that sample, though this conclusion is tenuous due to the substantial uncertainties ($30.0^{+20.8}_{-15.8}\%$ below $5$ kpc) and $1\sigma$ overlap with the offset AGN and \sdss~dual AGN.  

\subsection{Connection Between Nuclear Separation and AGN Luminosity}
\label{sec:sep_lum}

\begin{figure}
\hspace*{-0.2in} \includegraphics[width=3.3in]{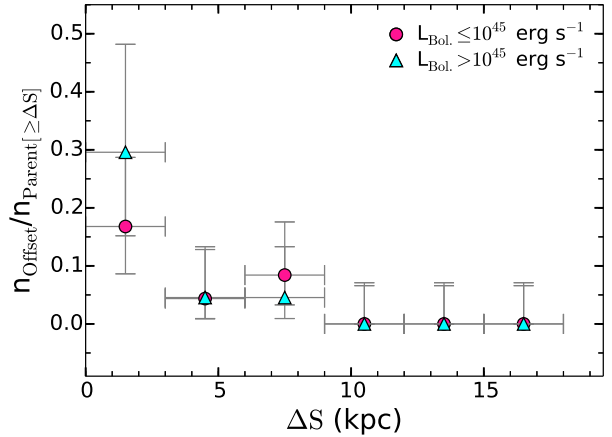}
\vspace*{-0.in}
\caption{\footnotesize{\offsetfrac~as a function \deltas~for the low-\lbol~subsample (pink circles) and high-\lbol~subsample (cyan triangles).  Vertical error bars correspond to $1\sigma$ binomial uncertainties, and horizontal errors bars denote the bin width.  Note that the two samples diverge by $\sim1\sigma$ at the smallest separation.}
}
\label{fig:frac_phys_sep_Lbollohi}
\end{figure}

Since the frequency of AGN observability in mergers peaks at small separations due to the increased availability of fuel for accretion, we may also expect that this regime coincides with enhanced accretion rates among AGN.  To investigate this prediction, we have examined the dependence of the offset AGN fractions as a function of nuclear separation for two subsamples that are separated by bolometric luminosity: \lbol~$\leq10^{45}$ erg s$^{-1}$ and  \lbol~$>10^{45}$ erg s$^{-1}$ (Figure \ref{fig:frac_phys_sep_Lbollohi}).  Due to the smaller numbers in each subsample compared to the full offset AGN sample, we have used bin sizes of $3$ kpc.

As shown in Figure \ref{fig:frac_phys_sep_Lbollohi}, the fractions in the low- and high-luminosity bins are consistent within their uncertainties throughout the entire range of separations investigated.  Therefore, we see no statistically significant evidence that the occurrence of offset AGN is dependent on luminosity at any given nuclear separation.  However, we note that the greatest divergence between the two fractions (by $\sim1\sigma$) occurs at the smallest separation.  While this result is tenuous due to the large uncertainties, it may hint that while an overall dependence of the offset AGN fraction with bolometric luminosity is not seen (Section \ref{sec:lbol_frac}), it does appear at small nuclear separations when the supply of gas for accretion is larger\footnotemark[7].  This result is also qualitatively consistent with the results from \citet{Koss:2012} in which the luminosities of dual AGN increase at smaller separations.  We caution that the observability of AGN in these systems depends not only on luminosity but also on the timescale of activity which we can not measure.  Still, these combined results are overall consistent with the numerical predictions that the bolometric luminosities of AGN in merging systems peak at merger stages corresponding to small separations \citep{Van_Wassenhove:2012,Stickley:2014}.

\footnotetext[7]{While this effect of luminosity could potentially be explained as a selection effect due to a tendency to find more offset AGN at smaller separations if they are brighter, such a selection would only exist if \lbol~was anti-correlated with \deltathetasig, which is not the case.}

\subsection{Nuclear Obscuration}
\label{sec:obscuration}

Heavy obscuration of the nuclear regions of galaxies undergoing mergers is predicted by models of galaxy and quasar co-evolution in which mergers trigger enhanced levels of accretion onto AGN but also pass through a stage of enhanced obscuration \citep{Hopkins2008}.  Therefore, the continuum emission of AGN that are hosted by on-going galaxy mergers may preferentially exhibit signs of obscuration compared to AGN in non-merging systems.  Extreme examples of AGN obscuration are found in ULIRGS that are rich in gas and dust \citep{Teng:2005}, with most emission from the accretion disk obscured except for high energy photons such as hard X-rays.  Heavily obscured (Compton-thick) AGN, in which X-ray emission is severely obscured, may constitute a significant population of AGN in galaxy mergers \citep{Kocevski:2015,Ricci:2017} and therefore understanding the role that X-ray absorption plays in the link between galaxy mergers and AGN is crucial.

The effect of obscuration in optically-selected and X-ray detected dual or offset AGN was noticed independently by \citet{Liu:2013} and \citet{Comerford:2015}.  Their samples of dual and offset Type 2 AGN systems show systematically lower observed hard X-ray to \oiii~luminosity ratios compared to the optically-selected Type 2 AGN sample from \citet{Heckman:2005}.  Since the \oiii~emission line originates far enough from the SMBH to not be subject to nuclear obscuration, this result suggests that dual and offset AGN are suffering heavier nuclear obscuration than the general population of Type 2 AGN.  In fact, \citet{Liu:2013} have shown that even the absorption-corrected hard X-ray luminosities of their sample are still under-luminous compared to a general sample of Type 2 AGN that was cross-matched with the $Chandra$ Source Catalogue \citep{Trichas:2012,Liu:2013}.  This has led to the suggestion that the low counts and/or intrinsically high absorbing columns result in systematically underestimated column densities. 

The analysis in Section \ref{sec:LOIII_LX} showed that, while the offset AGN sample has a lower mean hard X-ray to \oiii~luminosity ratio than the parent AGN, the difference is not at a significant level.  Therefore, no evidence of preferential nuclear obscuration is seen.  This result is opposite that seen in \citet{Liu:2013} and \citet{Comerford:2015}, and may be a result of the spatially offset selection introducing a bias toward face-on systems (as numerically predicted in \paperI) with shallower absorbing columns.  However, it may also be a result of the X-ray selection properties that target X-ray bright AGN with intrinsically little nuclear obscuration.  That the selection may preferentially target galaxy mergers with relatively dust-free nuclear regions means the sample is fundamentally different from the prototypical mergers seen in gas-rich systems such as ULIRGs and may be probing a specific subclass of galaxy mergers. 

\section{Conclusions}
\label{sec:conclusions}

We have used our systematically constructed sample of spatially offset AGN from \paperI~to constrain the parameters under which AGN triggering is driven by galaxy mergers.  Due to the selection of galaxy mergers based on offset X-ray AGN, our sample is not biased toward morphological disturbances or large projected physical separations, allowing us to investigate the AGN merger fraction in major or minor mergers and at early or late merger stages.  We have investigated the fractions of offset AGN, and those of similarly constructed dual AGN samples, out of their respective parent samples, as functions of AGN bolometric luminosity and projected nuclear separation.  Additionally, we have examined their group environments and compared their X-ray to optical luminosity ratios to those of independent AGN samples.  Our conclusions are as follows:

\begin{enumerate}

\item The fraction of spatially offset AGN shows no evidence for a dependence on AGN bolometric luminosity, while the fractions of dual AGN do show a positive dependence, increasing from $0\%$ at $10^{42}$ erg s$^{-1}$ to between $\sim10\%$ and $\sim40\%$ at $10^{46}$ erg s$^{-1}$ (Figure \ref{fig:frac_Lbol_OIII}).  These results suggest that AGN triggering is indeed linked to mergers but that this dependence may only become strong in the specific scenarios of high bolometric luminosities, dual AGN activation and possibly major mergers.

\item The offset AGN group environments show no evidence for a difference from the parent AGN sample (Figure \ref{fig:hist_ngroup}), and both reside in environments with a low-density of massive galaxies.  The lack of numerous massive companions may point toward a preference for minor mergers in the offset AGN systems.

\item The fractions of spatially offset AGN and dual AGN show evidence for a negative dependence on projected physical nuclear separation, increasing from $0\%$ at 19 kpc to between $\sim5\%$ and $30\%$ at $<3$ kpc (Figure \ref{fig:frac_phys_sep}).  The offset and dual AGN fractions are similar at small separations, suggesting that the efficiency of dual AGN triggering becomes similar to single AGN triggering at late merger stages when significant material is available for accretion.  We can only trace the dual AGN sample down to $\sim2$ kpc, while the resolution of our offset AGN sample allows it to be traced down to $\sim0.8$ kpc where we see the most significant increase.  Our sample of offset AGN has allowed this analysis to be extended down to $<1$ kpc for the first time.   

\item We see tentative evidence that the inverse dependence of the AGN merger fractions on separation become strongest when restricted to a high AGN luminosity subsample (Figure \ref{fig:frac_phys_sep_Lbollohi}).  If real, this result would be consistent with numerical predictions that AGN triggering probabilities increase with decreasing nuclear separations, and that this late merger stage also corresponds with the stage of highest AGN luminosity.

\item The hard X-ray to \oiii~luminosity ratios of offset AGN show no significant evidence for a difference from that of the parent AGN (Figure \ref{fig:plot_LX_LOIII}), in contrast to the expectation from studies of many known merging galaxy systems.  While this similarity may reflect a tendency to select face-on systems, it may also point toward a selection of intrinsically unobscured systems that are fundamentally different from merging systems with coincident nuclear obscuration and on-going star-formation.

\end{enumerate}

In Paper III of this series, we will present new and archival $HST$ imaging for a subset of our offset AGN sample to put constraints on the correlated evolution of SMBHs and their host galaxies.  In particular, we will determine the effect of merger mass ratio on SMBH growth, and we will put constraints on the correlated triggering of star-formation and AGN.  \\

The authors would like to thank an anonymous referee for detailed and insightful comments that greatly improved the quality of the paper.  Support for this work was provided by NASA through Chandra Award Number AR5-16010A issued by the Chandra X-ray Observatory Center, which is operated by the Smithsonian Astrophysical Observatory for and on behalf of NASA under contract NAS8-03060. The scientific results reported in this article are based in part on observations made by the \emph{Chandra X-ray Observatory}, and this research has made use of data obtained from the \emph{Chandra Data Archive} and the \emph{Chandra Source Catalog}, and software provided by the \emph{Chandra X-ray Center (CXC)} in the application packages \emph{CIAO}, \emph{ChIPS}, and \emph{Sherpa}.

\end{document}